\documentclass[a4paper]{article}
\usepackage{INTERSPEECH2021}
\usepackage{enumitem}
\usepackage{multirow}
\usepackage{xspace}

\newcommand{\ADReSSo}{ADReSS$_o$\xspace}

 \title{Detecting cognitive decline using speech only: \\The \ADReSSo Challenge}
\name{Saturnino Luz$^1$, Fasih Haider$^1$, Sofia de la Fuente$^1$, Davida Fromm$^2$,  Brian MacWhinney$^2$}
\address{
  $^1$Usher Institute, Edinburgh Medical School, The University of
  Edinburgh, UK\\
  $^2$Department of Psychology, Carnegie Mellon University, USA}
\email{\{S.Luz, fasih.haider, sofia.delafuente\}@ed.ac.uk, \{fromm, macw\}@andrew.cmu.edu}

\begin{document}

\maketitle
\begin{abstract}
Building on the success of the ADReSS Challenge at Interspeech 2020, which attracted the participation of 34 teams from across the world, the \ADReSSo Challenge targets three difficult automatic prediction problems of societal and medical relevance, namely: detection of Alzheimer's Dementia, inference of cognitive testing scores, and prediction of cognitive decline. This paper presents these prediction tasks in detail, describes the datasets used, and reports the results of the baseline classification and regression models we developed for each task. A combination of acoustic and linguistic features extracted directly from audio recordings, without human intervention, yielded a baseline accuracy of 78.87\% for the AD classification task, an MMSE prediction root mean squared (RMSE) error of 5.28, and 68.75\% accuracy for the cognitive decline prediction task.
\end{abstract}

\section{Introduction}

Dementia is a category of neurodegenerative diseases which entail
long-term and usually gradual decrease of cognitive functioning.  The
main risk factor for dementia is age and, hence, it is increasingly
prevalent in our ageing society. Due to the severity of the disease,
institutions and researchers worldwide are investing considerably on
dementia prevention, early detection and disease progression
monitoring \cite{bib:RitchieCarriereEtAl17al}.  There is a need for
cost-effective and scalable methods for detection of early signs of
Alzheimer's Dementia (AD) as well as prediction of disease
progression.

Methods for screening and tracking the progression of dementia
traditionally involve cognitive tests such as the Mini-Mental Status
Examination (MMSE) \cite{folstein1975mini} and the Montreal Cognitive
Assessment (MoCA) \cite{nasreddine2005montreal}. MMSE and MoCA are
widely used because, unlike imaging methods, they are cheap, quick to
administer and easy to score. Despite its shortcomings in specificity
in early stages of dementia, the MMSE is still widely used.  The
promise of speech technology in comparison to cognitive tests is
twofold. First, speech can be collected passively, naturally and
continuously throughout the day, gathering increasing data points
while burdening neither the participant nor the researcher.
Furthermore, the combination of speech technology and machine learning
creates opportunities for automatic screening systems and diagnosis
support tools for dementia. The \ADReSSo Challenge aims to generate
systematic evidence for these promises towards their clinical
implementation.


As with the its predecessor, the overall objective of the \ADReSSo
Challenge is to host a shared task for the systematic comparison of
approaches to the detection of cognitive impairment and decline based
on spontaneous speech. As has been pointed out elsewhere
\cite{bib:LuzHaiderEtAl20ADReSS,bib:DelaFuenteRichieLuz2020JAD}, the
lack of common, standardised datasets and tasks has hindered the
benchmarking of the various approaches proposed to date, resulting in
a lack of translation of these speech based methods into clinical
practice. 

The \ADReSSo Challenge provides a forum for researchers working on
approaches to cognitive decline detection based on speech data to test
their existing methods or develop novel approaches on a new shared
standardised dataset. The approaches that performed best on last
year's dataset \cite{bib:LuzHaiderEtAl20ADReSS} employed features
extracted from manual transcripts which were provided along with the
audio data
\cite{bib:SyedSyedEtAl20autsald,bib:YuanBianEtAl20dftp}. The best
performing method \cite{bib:YuanBianEtAl20dftp} made interesting use
of pause and disfluency annotation provided with the
transcripts. While this provided interesting insights into the
predictive power of these paralinguistic features for detection of
cognitive decline, extracting such features, and indeed accurate
transcripts from spontaneous speech remains an open research
issue.  This year's \ADReSSo (Alzheimer's Dementia Recognition
through Spontaneous Speech {\em only}) tasks provide more challenging
and improved spontaneous speech datasets, requiring the creation of
models straight from speech, without manual transcription, though
automatic transcription is allowed and encouraged.

The \ADReSSo datasets are carefully matched so as to avoid 
common biases often overlooked in evaluations of AD detection methods,
including repeated occurrences of speech from the same participant
(common in longitudinal datasets), variations in audio quality, and
imbalances of gender and age distribution.
The challenge defines three tasks:   
\begin{enumerate}
\item an AD classification task, where participants were required to produce a
  model to predict the label (AD or non-AD) for a short speech
  session. Participants 
  could use the speech signal directly (acoustic features), or attempt
  to convert the speech into text automatically (ASR) and extract
  linguistic features from this automatically generated transcript; 
\item an MMSE score regression task, where participants were asked to
  create models to infer the patients' MMSE score based solely on speech
  data; and
\item a cognitive decline (disease progression) inference task, where
  they created models for prediction of changes in cognitive status
  over time, for a given speaker, based on speech data collected at
  baseline (i.e. the beginning of a cohort study).
\end{enumerate}

These tasks depart from neuropsychological and clinical
evaluation approaches that have employed speech and language
\cite{bib:TalerPhillips08lal} by focusing on prediction recognition
using spontaneous speech.  Spontaneous speech analysis has the
potential to enable novel applications for speech technology in
longitudinal, unobtrusive monitoring of cognitive health
\cite{bib:LuzCBMS17}, in line with the theme of this year's
INTERSPEECH, ``Speech Everywhere!''.

This paper describes the \ADReSSo dataset and presents baselines for
all \ADReSSo tasks, including feature extraction procedures and models
for AD detection, MMSE score regression and prediction of cognitive
decline.

\section{Related work}
\label{sec:related-work}

There has been increasing research on speech technology for dementia
detection over the last decade. The majority of this research has
focused on AD classification, but some of it targets MCI detection as
well \cite{bib:DelaFuenteRichieLuz2020JAD}. These objectives are most
closely related with our first task, namely, the AD classification
task. Such related research includes the best performing models
presented in the ADReSS challenge in 2020. These achieved an 85.45\%
\cite{bib:SyedSyedEtAl20autsald} and 89.6\%
\cite{bib:YuanBianEtAl20dftp} accuracy in AD classification using
acoustic features and text-based features extracted from manual
transcripts. Classification based on acoustic features only was also
attempted in \cite{bib:SyedSyedEtAl20autsald}, and obtained 76.85\%
accuracy with IS10-Paralinguistics feature set (a low dimensional
version of ComParE \cite{bib:SchullerSteidlEtAl10in}) and
Bag-of-Acoustic-Words (BoAW).

Few works rely exclusively on acoustic features or text features
extracted through ASR. One of these achieved a 78.7\% accuracy on a
subset of the Cookie Theft task of the Pitt dataset, using different
comprehensive paralinguistic feature sets and standard machine
learning algorithms \cite{bib:HaiderFuenteLuz20aspacf}. Another, using the
complete Pitt dataset, obtained 68\% accuracy using only vocalisation
features (i.e. speech-silence patterns)
\cite{bib:LuzCBMS17}. Classification accuracy of 62.3\% has been
reported for a different spontaneous speech dataset using fully
automated ASR features \cite{mirheidari2018detecting}.

As regards the second task, regression over MMSE scores, there is less
literature available and most of it has been presented in recent
workshops \cite{bib:DelaFuenteRichieLuz2020JAD}. Several of these
works used the above mentioned Pitt dataset to extract different
linguistic and acoustic features and predict MMSE scores. A recent
study captured different levels of cognitive impairment with a
multiview embedding and obtained a mean absolute error (MAE) of 3.42
\cite{pou2018learning}. Another study reported a MAE of 3.1, having
relied solely on acoustic features to build their regression model (a
set of 811 features) \cite{al2017detecting}. Error scores as low as
2.2 (MAE) have been obtained, but relying on non-spontaneous speech
data such as elicited in semantic verbal fluency (SVF) tasks
\cite{linz2017predicting}.

Studies addressing the progression task are far less common. Notable
in this category is \cite{yancheva2015using}, which incorporated a
comprehensive set of features (i.e. lexicosyntactic, semantic and
acoustic) into Bayesian network with, reporting a MAE of 3.83 on
prediction of MMSE scores throughout different study visits.  Two
other studies account for disease progression in classification
experiments. One of them
extracted speech-based from the ISLE dataset achieving 80.4\% accuracy to detect
intra-subject cognitive changes, that is, to distinguish healthy
participants who remained healthy from those who developed some kind
of cognitive impairment \cite{Weiner2016}. The second study uses SVF
scores to build a machine learning classifier able to predict changes
from MCI to AD over a 4-year follow-up, with 84.1\% accuracy
\cite{Clark2016}.

\section{The \ADReSSo  Datasets}
\label{sec:adress-chall-dataset}

We provided two distinct datasets for the \ADReSSo Challenge:

\begin{enumerate}
\item a dataset consisting of speech recordings of Alzheimer's patients
  performing a category (semantic) fluency task \cite{bib:Benton68d}
  at their baseline visit, for prediction of cognitive decline over a
  two year period, and
\item a set of recordings of picture descriptions produced by
  cognitively normal subjects and patients with an AD diagnosis, who
  were asked to describe the Cookie Theft picture from the Boston
  Diagnostic Aphasia Exam \cite{Becker1994,bib:GoodglassKaplanBarresi01b}.
\end{enumerate}

The recorded data also included speech from different experimenters
who gave instructions to the patients and occasionally interacted with
them in short dialogues. No transcripts were provided with either
dataset, but segmentations of the recordings into vocalisation
sequences with speaker identifiers \cite{bib:LuzFuenteAlbert18manps}
were made available for optional use. The \ADReSSo challenge's
participants were asked to specify whether they made use of these
segmentation profiles in their predictive modelling.  Recordings were
acoustically enhanced with stationary noise removal and audio volume
normalisation was applied across all speech segments to control for
variation caused by recording conditions such as microphone placement.

\begin{figure}[htb]
  \centering
  \includegraphics[width=\linewidth]{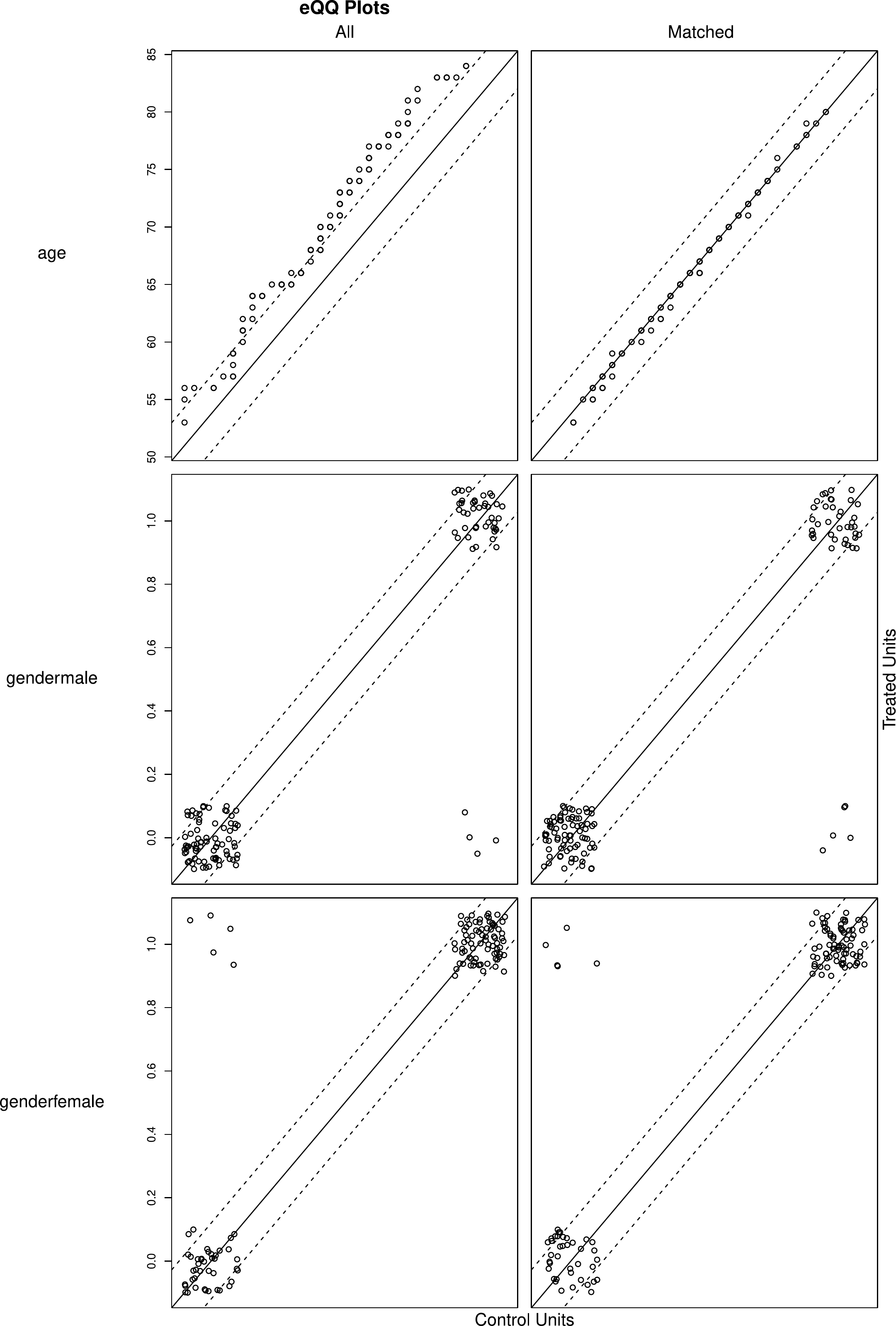}
  \caption{Quantile-quantile plots for data before (left) and after 
    matching (right) by age and gender. }
  \label{fig:qqplots}
\end{figure}

The dataset used for AD and MMSE was matched for age and gender so as
to minimise risk of bias in the prediction tasks. We matched the data
using a propensity score approach
\cite{bib:RosenbaumRubin83,bib:Rubin73mrbob} implemented in the R
package MatchIt \cite{bib:HoImaiEtAl11m}. The final dataset matched
according to propensity scores defined in terms of the probability of
an instance of being treated as AD given covariates age and gender.
All standardised mean differences for the age and gender covariates
were $< 0.001$ and all standardised mean differences for
$\text{age}^2$ and two-way interactions between covariates were well
below .1, indicating adequate balance for the covariates. The
propensity score was estimated using a probit regression of the
treatment on the covariates age and gender as probit generated a
better balanced than logistic regression. The age/gender matching is
summarised in Figure~\ref{fig:qqplots}, which shows the respective
(empirical) quantile-quantile plots for the original and balanced
datasets. As usual, a quantile-quantile plot showing instances near
the diagonal indicates good balance.

The resulting dataset encompasses
242 audio files. These were split into training and test sets, with
70\% of instances allocated to the former and 30\% allocated to the
latter. These partitions were generated so as to preserve gender and
age matching. The overall characteristics of this dataset are shown in
Table~\ref{tab:dx}. 

\begin{table}[h]
\centering
\caption{Characteristics of patients in the Diagnostic tasks dataset.}\vspace{-1ex}
\label{tab:dx}
\begin{tabular}[htb]{ccc}
  \hline
  TO BE ADDED \\
  \hline
  \\
  \\
  \hline
\end{tabular}
\end{table}

The dataset for the disease prognostics task (prediction of cognitive
decline) was created from a longitudinal cohort study involving AD
patients. The time period for assessment of disease progression
spanned the baseline and the year-2 data collection visits of the
patients to the clinic. The task involves classifying patients into
'decline' or 'no-decline' categories, given speech collected at
baseline as part of a category fluency test. Decline was defined as a
difference in MMSE score between baseline and year-2 greater than or
equal 5 points (i.e. $mmse(baseline) - mmse(y2) \ge 5$). This dataset
has a total of 105 audio recordings split into training and test sets
as with the diagnosis dataset (70\% and 30\% of recordings,
respectively). The dataset is summarise in Table~\ref{tab:px}

\begin{table}[h]
\centering
\caption{Characteristics of patients in the prognostic tasks dataset.}\vspace{-1ex}
\label{tab:px}
\begin{tabular}[htb]{ccc}
  \hline
  TO BE ADDED \\
  \hline
  \\
  \\
  \hline
\end{tabular}
\end{table}

\section{Data representation}
\label{sec:data-representation}

\subsection{Acoustic features}
\label{sec:acoustic-features}

We applied a sliding window with a length of 100 ms on the audio files
of the dataset with no overlap and extracted \textit{eGeMAPS} features
over such frames. The \textit{eGeMAPS} feature set
\cite{bib:EybenSchererEtAl16gg} resulted from an attempt to reduce the
somewhat unwieldy feature sets above to a basic set of acoustic
features based on their potential to detect physiological changes in
voice production, as well as theoretical significance and proven
usefulness in previous studies. It contains the F0 semitone, loudness,
spectral flux, MFCC, jitter, shimmer, F1, F2, F3, alpha ratio,
Hammarberg index and slope V0 features, as well as their most common
statistical functionals, totalling 88 features per 100ms frame. We
then applied the active data representation method (ADR)
\cite{bib:HaiderFuenteLuz20aspacf} to generate a data representation
using frame level acoustic information for each audio recording. The
ADR method has been tested previously for generating representations
for large scale time-series data. It employs self-organising mapping
to cluster the original acoustic features and then computes
second-order features over these cluster to extract new features (see
\cite{bib:HaiderFuenteLuz20aspacf} for details). Note that this method
is entirely automatic in that no speech segmentation of diarisation
information is provided to the algorithm.

\subsection{Linguistic Features}

We used the Google Cloud-based Speech Recogniser for automatically
transcribing the audio files. The transcripts were converted into CHAT
format which is compatible with CLAN \cite{bib:MacWhinney17t}, a set
of programs that allows for automatic analysis of a wide range of
linguistic and discourse structures. Next, we used the automated MOR
function to assign lexical and morphological descriptions to all the
words in the transcripts. Then, we used two commands: EVAL which
creates a composite profile of 34 measures, and FREQ to compute the
Moving Average Type Token Ratio \cite{covington2010cutting}.

\section{Diagnosis baseline}
\label{sec:diagn-basel-syst}

\subsection{Task 1: AD Classification}

The AD classification experiments were performed using five different
methods, namely decision trees (DT, where the~leaf size is optimised
through a~grid search within a~range of 1 to 20), nearest neighbour
(KNN, where K parameter is optimised through a~grid search within
a~range of 1 to 20), linear discriminant analysis (LDA), Tree Bagger
(TB, with 50 trees, where leaf size is optimised through a~grid search
within a~range of 1 to 20), and~support vector machines (SVM, with
a~linear kernel, where box constraint is optimised by trying a~grid
search between 0.1 to 1.0), and sequential minimal optimisation
solver.

\begin{figure}[htb]
  \centering
  \includegraphics[width=.7\linewidth]{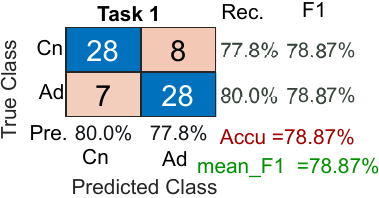}
  \caption{Task 1: Late (decision) fusion of the best results of acoustic and linguistic models. Precision (Pre) , recall (Rec), accuracy (Accu) and mean $F_1$ scores are shown on the margins.}
  \label{fig:task1matrix}
\end{figure}

The results for accuracy in the AD vs Control (CN) classification task
are summarised in Table~\ref{tab:task1results}. As indicated in
boldface, the best performing classifier in cross-validation was DT,
achieving 78.92\% and 72.89\% accuracy using acoustic and linguistic
features, respectively. On the test set, however, the results were
reversed, with linguistic features producing an overall best accuracy
of 77.46\%, with the SVM classifier. Late fusion of the acoustic and linguistic models improves the accuracy on the test set further to 78.87\% (Figure~\ref{fig:task1matrix}). 

\begin{table}[htbp]
  \caption{Task1: AD classification accuracy on leave-one-subject out cross-validation (CV) and test data.}
  \label{tab:task1results}
  \vspace{-1ex}\small
\begin{tabular}{@{}l@{~~}c@{~~~}c@{~~}c@{~~}c@{~~}c@{~~}c@{~~}c@{}}
\hline 
                      & Classifier & LDA   & DT          & SVM            & RF    & KNN   & mean (sd)   \\\hline
\multirow{2}{*}{CV}   & Acoustic   & 62.65 & {\bf 78.92} & 69.28          & 65.06 & 65.06 & 68.19 (6.4) \\
                      & Linguistic & 72.29 & {\bf 72.89} & 72.89          & 75.90 & 65.06 & 71.81 (4.0)  \\                  
                  \hline
\multirow{2}{*}{Test} & Acoustic   & 50.70 & 60.56       & \textbf{64.79} & 63.38 & 53.52 & 58.59 (6.2)  \\
                      & Linguistic & 76.06 & 74.65       & \textbf{77.46} & 73.24 & 59.15 & 72.11 (7.4)  \\\hline
\end{tabular}
\end{table}

\subsection{Task 2: MMSE prediction}
\label{sec:mmse-prediction-models}

For this regression task we also used five types of regression models:
linear regression (LR), DT, with leaf size of 20 and
CART algorithm, support vector regression (SVR, with a radial basis
function kernel with box constraint of 0.1, and sequential minimal
optimisation solver), Random Forest regression ensembles (RF),
and Gaussian process regression (GP, with a squared exponential
kernel).  The regression methods are implemented in MATLAB
\cite{bib:MATLAB19} using the statistics and machine learning toolbox.

\begin{table}[htbp]
  \caption{Task2: MMSE score prediction error scores (RMSE) for leave-one-subject out CV and test data.}
  \label{tab:task2results}\small\vspace{-1ex}
  \begin{tabular}{@{}l@{~~}c@{~~~~}c@{~~~}c@{~~~}c@{~~~}c@{~~~}c@{~~~}c@{}}
    \hline 
                      & Regression & LR   & DT   & SVR           & RF   & GP   & mean (sd)   \\\hline
\multirow{3}{*}{CV}   & Acoustic   & 6.88 & 6.88 & 6.96          & 7.89 & 6.71 & 7.06 (0.47) \\
                      & Linguistic & 6.65 & 5.92 & 6.42          & 7.02 & 6.50 & 6.50 (0.40) \\
                  \hline
\multirow{3}{*}{Test} & Acoustic   & 6.23 & 6.47 & \textbf{6.09} & 8.18 & 6.81 & 6.75 (0.84) \\
                      & Linguistic & 5.87 & 6.24 & \textbf{5.28} & 6.94 & 5.43 & 5.95 (0.67) \\\hline    
\end{tabular}
\end{table}

The results are summarised in Table~\ref{tab:task2results}. As with
classification, DT regression outperformed the other models in
cross-validation, with ASR linguistic features outperforming acoustic
ADR features. This trend persisted in the test set, with linguistic
features producing a minimum RMSE of 5.28 in a SVR model.
We then fused the best results of linguistic and acoustic features and took
a weighted mean, finding the weights through grid search on the
validation results, which resulted in an improvement (6.37) on the
validation dataset. We then used the same weights to fuse the test
results and obtained an RMSE of 5.29 ($r = 0.69$).

\section{Prognosis baseline}
\label{sec:prognosis-baseline}

\subsection{Task 3: prediction of progression}

We tested the same classification methods used in Task 1 for the task
of identifying those patients who went on to exhibit cognitive decline
within two years of the baseline visit in which the speech samples
used in our models were taken. The acoustic and linguistic features
were generated as described in
Section~\ref{sec:data-representation}.The results of this prediction
task are summarised in Table~\ref{tab:task3results}. As the classes
for this task are imbalanced we report average $F_1$ results rather
than accuracy, Once again DT performed best on CV, but the $F_1$
results for the test set was considerably lower, reaching only a
maximum of 66.67\%, for linguistic features and 61.02\% for acoustic
features.

\begin{table}[htbp]
  \caption{Task3: cognitive decline progression results (mean of
    $F_{1}Score$) for leave-one-subject-out CV and test data.}\label{tab:task3results}
  \small
  \begin{tabular}{@{}l@{~~}c@{~~~}c@{~~}c@{~~}c@{~~}c@{~~}c@{~~}c@{}}
  \hline 
                      & Classifier & LDA         & DT         & SVM   & RF    & KNN   & mean (sd)     \\\hline
\multirow{3}{*}{Val}  & Acoustic   & 59.89       & 84.94      & 55.64 & 63.85 & 65.92 & 66.05 (11.27) \\
                      & Linguistic & 55.19       & 76.52      & 45.24 & 63.10 & 55.25 & 59.06 (11.64) \\
                   \hline
\multirow{3}{*}{test} & Acoustic   & {\bf 61.02} & 53.62      & 40.74 & 40.74 & 38.46 & 46.91 (9.89)  \\
                      & Linguistic & 54.29       & {\bf 66.67} & 40.74 & 56.56 & 39.62 & 51.58 (11.41) \\\hline
\end{tabular}
\end{table}

\begin{figure}[htb]
  \centering
  \includegraphics[width=.6\linewidth]{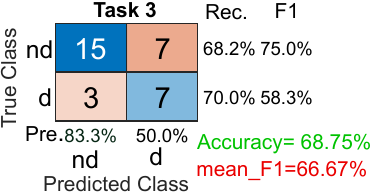}
  \caption{Task 3: Decision fusion of the best results of acoustic and
    linguistic features on the test set.}
  \label{fig:task3matrix}
\end{figure}

As before, we fused the predictions of the best models for each
feature type, hoping that the diversity of models might improve
classification. The confusion matrix for the fusion model is shown in
Figure~\ref{fig:task3matrix}. This time, however, decision fusion did
not yield any improvement.

\section{Discussion}

The AD classification baseline yielded a maximum accuracy of 78.87\%
on the test set, through the fusion of models based on linguistic and
acoustic features. Despite the fact that the ASR transcripts had
relatively high word error rates, linguistic features contributed
considerably to the predictions. The overall baseline results for this
task are in fact comparable to results obtained for similar picture
description data using manual transcripts (see
Section~\ref{sec:related-work}). DT classifiers performed well on the
CV experiments, but accuracy decreased on the test set, indicating
probable overfitting. Overall, however, all models proved fairly
robust.

A similar picture was observed in the MMSE regression
task. Linguistic features contributed appreciably to the prediction,
even though the transcripts contained many errors. In this case,
however, late fusion only improved the RMSE score in CV; the test set
RMSE remained practically unchanged.

The prognosis task proved to be the most difficult prediction
task. The CV results varied considerably among models, with a standard
deviation of 11.64 for the linguistic models. The test set results were
also varied, reaching a maximum $F_1$ score of 66.67\%, even when the
best model predictions were fused. Although the acoustic features
produced the best classification results in CV ($F_1 = 66.05\%$ vs
59.06\% for linguistic features), these results were not born out by
test set evaluation, suggesting that the acoustic features made the
classifiers more prone to overfitting. It is possible that this could
be mitigated by training the acoustic feature extractor (ADR) on a
larger set of recordings (data augmentation) and fine tuning the
resulting model on the \ADReSSo data.

\section{Conclusions}

The \ADReSSo Challenge is the first shared task to target cognitive
status prediction using raw, non-annotated a non-transcribed speech,
and to address prediction of changes in cognition over time. We
believe this moves the speech processing and machine learning methods
one step closer to the real-world of clinical applications. A
limitation the AD classification and the MMSE regression tasks share
with most approaches to the use of these methods in dementia research
is that they provide little insight into disease progression. This has
been identified as the main issue hindering translation of these
technologies into clinical practice
\cite{bib:DelaFuenteRichieLuz2020JAD}. However, these tasks remain
relevant in application scenarios involving automatic cognitive status
monitoring, in combination with wearable and ambient technology. The
addition of the progression task should open avenues for relevance
also in more traditional clinical contexts.

\section{Acknowledgements}
This research is funded by the European Union's Horizon 2020 research
programme, under grant agreement 769661, SAAM project.
SdlFG is supported by the Medical Research Council.

\bibliographystyle{IEEEtran}
\bibliography{IEEEabrv,adrsopaper}

\begin{thebibliography}{10}
\providecommand{\url}[1]{#1}
\csname url@samestyle\endcsname
\providecommand{\newblock}{\relax}
\providecommand{\bibinfo}[2]{#2}
\providecommand{\BIBentrySTDinterwordspacing}{\spaceskip=0pt\relax}
\providecommand{\BIBentryALTinterwordstretchfactor}{4}
\providecommand{\BIBentryALTinterwordspacing}{\spaceskip=\fontdimen2\font plus
\BIBentryALTinterwordstretchfactor\fontdimen3\font minus
  \fontdimen4\font\relax}
\providecommand{\BIBforeignlanguage}[2]{{%
\expandafter\ifx\csname l@#1\endcsname\relax
\typeout{** WARNING: IEEEtran.bst: No hyphenation pattern has been}%
\typeout{** loaded for the language `#1'. Using the pattern for}%
\typeout{** the default language instead.}%
\else
\language=\csname l@#1\endcsname
\fi
#2}}
\providecommand{\BIBdecl}{\relax}
\BIBdecl

\bibitem{bib:RitchieCarriereEtAl17al}
K.~Ritchie, I.~Carrière, L.~Su, J.~T. O'Brien, S.~Lovestone, K.~Wells, and
  C.~W. Ritchie, ``The midlife cognitive profiles of adults at high risk of
  late-onset {Alzheimer}'s disease: The {PREVENT} study,'' \emph{Alzheimer's \&
  Dementia}, vol.~13, no.~10, pp. 1089--1097, 2017.

\bibitem{folstein1975mini}
M.~F. Folstein, S.~E. Folstein, and P.~R. McHugh, ````mini-mental state'': a
  practical method for grading the cognitive state of patients for the
  clinician,'' \emph{Journal of psychiatric research}, vol.~12, no.~3, pp.
  189--198, 1975.

\bibitem{nasreddine2005montreal}
Z.~S. Nasreddine, N.~A. Phillips, V.~B{\'e}dirian, S.~Charbonneau,
  V.~Whitehead, I.~Collin, J.~L. Cummings, and H.~Chertkow, ``The montreal
  cognitive assessment, moca: a brief screening tool for mild cognitive
  impairment,'' \emph{Journal of the American Geriatrics Society}, vol.~53,
  no.~4, pp. 695--699, 2005.

\bibitem{bib:LuzHaiderEtAl20ADReSS}
\BIBentryALTinterwordspacing
S.~Luz, F.~Haider, S.~de~la Fuente, D.~Fromm, and B.~MacWhinney,
  ``{Alzheimer's} dementia recognition through spontaneous speech: The {ADReSS
  Challenge},'' in \emph{Proceedings of {INTERSPEECH 2020}}, Shanghai, China,
  2020. [Online]. Available: \url{https://arxiv.org/abs/2004.06833}
\BIBentrySTDinterwordspacing

\bibitem{bib:DelaFuenteRichieLuz2020JAD}
S.~de~la Fuente~Garcia, C.~Ritchie, and S.~Luz, ``Artificial intelligence,
  speech and language processing approaches to monitoring {Alzheimer's}
  disease: a systematic review,'' \emph{Journal of Alzheimer's Disease},
  vol.~78, no.~4, pp. 1547--1574, 2020.

\bibitem{bib:SyedSyedEtAl20autsald}
M.~S.~S. Syed, Z.~S. Syed, M.~Lech, and E.~Pirogova, ``{Automated Screening for
  Alzheimer’s Dementia Through Spontaneous Speech},'' in \emph{Proc.
  Interspeech 2020}, 2020, pp. 2222--2226.

\bibitem{bib:YuanBianEtAl20dftp}
J.~Yuan, Y.~Bian, X.~Cai, J.~Huang, Z.~Ye, and K.~Church, ``{Disfluencies and
  Fine-Tuning Pre-Trained Language Models for Detection of Alzheimer’s
  Disease},'' in \emph{Proc. Interspeech 2020}, 2020, pp. 2162--2166.

\bibitem{bib:TalerPhillips08lal}
V.~Taler and N.~A. Phillips, ``Language performance in alzheimer's disease and
  mild cognitive impairment: A comparative review,'' \emph{Journal of Clinical
  and Experimental Neuropsychology}, vol.~30, no.~5, pp. 501--556, 2008.

\bibitem{bib:LuzCBMS17}
S.~Luz, ``Longitudinal monitoring and detection of {Alzheimer's} type dementia
  from spontaneous speech data,'' in \emph{Computer Based Medical
  Systems}.\hskip 1em plus 0.5em minus 0.4em\relax IEEE Press, 2017, pp.
  45--46.

\bibitem{bib:SchullerSteidlEtAl10in}
B.~Schuller, S.~Steidl, A.~Batliner, F.~Burkhardt, L.~Devillers, C.~M{\"u}ller,
  and S.~S. Narayanan, ``The {INTERSPEECH 2010} paralinguistic challenge,'' in
  \emph{Procs. of the 11th Annual Conference of the International Speech
  Communication Association, INTERSPEECH}, 2010, pp. 2794--2797.

\bibitem{bib:HaiderFuenteLuz20aspacf}
F.~Haider, S.~de~la Fuente, and S.~Luz, ``An assessment of paralinguistic
  acoustic features for detection of alzheimer's dementia in spontaneous
  speech,'' \emph{IEEE Journal of Selected Topics in Signal Processing},
  vol.~14, no.~2, pp. 272--281, 2020.

\bibitem{mirheidari2018detecting}
B.~Mirheidari, D.~Blackburn, T.~Walker, A.~Venneri, M.~Reuber, and
  H.~Christensen, ``{Detecting Signs of Dementia Using Word Vector
  Representations.}'' in \emph{Interspeech}, 2018, pp. 1893--1897.

\bibitem{pou2018learning}
C.~Pou-Prom and F.~Rudzicz, ``Learning multiview embeddings for assessing
  dementia,'' in \emph{Proceedings of the 2018 Conference on Empirical Methods
  in Natural Language Processing}, 2018, pp. 2812--2817.

\bibitem{al2017detecting}
S.~Al-Hameed, M.~Benaissa, and H.~Christensen, ``Detecting and predicting
  alzheimer's disease severity in longitudinal acoustic data,'' in
  \emph{Proceedings of the International Conference on Bioinformatics Research
  and Applications 2017}, 2017, pp. 57--61.

\bibitem{linz2017predicting}
N.~Linz, J.~Tr{\"o}ger, J.~Alexandersson, M.~Wolters, A.~K{\"o}nig, and
  P.~Robert, ``Predicting dementia screening and staging scores from semantic
  verbal fluency performance,'' in \emph{2017 IEEE International Conference on
  Data Mining Workshops (ICDMW)}.\hskip 1em plus 0.5em minus 0.4em\relax IEEE,
  2017, pp. 719--728.

\bibitem{yancheva2015using}
M.~Yancheva, K.~C. Fraser, and F.~Rudzicz, ``Using linguistic features
  longitudinally to predict clinical scores for alzheimer’s disease and
  related dementias,'' in \emph{Proceedings of SLPAT 2015: 6th Workshop on
  Speech and Language Processing for Assistive Technologies}, 2015, pp.
  134--139.

\bibitem{Weiner2016}
J.~Weiner and T.~Schultz, ``{Detection of Intra-Personal Development of
  Cognitive Impairment From Conversational Speech},'' in \emph{Speech
  Communication; 12. ITG Symposium}, 2016, pp. 1--5.

\bibitem{Clark2016}
D.~G. Clark, P.~M. McLaughlin, E.~Woo, K.~Hwang, S.~Hurtz, L.~Ramirez,
  J.~Eastman, R.~M. Dukes, P.~Kapur, T.~P. DeRamus, and L.~G. Apostolova,
  ``{Novel verbal fluency scores and structural brain imaging for prediction of
  cognitive outcome in mild cognitive impairment},'' \emph{Alzheimer's and
  Dementia: Diagnosis, Assessment and Disease Monitoring}, vol.~2, pp.
  113--122, 2016.

\bibitem{bib:Benton68d}
A.~L. Benton, ``Differential behavioral effects in frontal lobe disease,''
  \emph{Neuropsychologia}, vol.~6, no.~1, pp. 53--60, 1968.

\bibitem{Becker1994}
J.~T. Becker, F.~Boller, O.~L. Lopez, J.~Saxton, and K.~L. McGonigle, ``{The
  Natural History of {Alzheimer}'s Disease},'' \emph{Archives of Neurology},
  vol.~51, no.~6, p. 585, 1994.

\bibitem{bib:GoodglassKaplanBarresi01b}
H.~Goodglass, E.~Kaplan, and B.~Barresi, \emph{{BDAE-3}: {Boston Diagnostic
  Aphasia Examination} -- Third Edition}.\hskip 1em plus 0.5em minus
  0.4em\relax Lippincott Williams \& Wilkins Philadelphia, PA, 2001.

\bibitem{bib:LuzFuenteAlbert18manps}
S.~Luz, S.~de~la Fuente, and P.~Albert, ``A method for analysis of patient
  speech in dialogue for dementia detection,'' in \emph{Resources for
  processing of linguistic, paralinguistic and extra-linguistic data from
  people with various forms of cognitive impairment}, D.~Kokkinakis, Ed.\hskip
  1em plus 0.5em minus 0.4em\relax ELRA, May 2018, pp. 35--42.

\bibitem{bib:RosenbaumRubin83}
P.~R. Rosenbaum and D.~B. Rubin, ``{The central role of the propensity score in
  observational studies for causal effects},'' \emph{Biometrika}, vol.~70,
  no.~1, pp. 41--55, 04 1983.

\bibitem{bib:Rubin73mrbob}
D.~B. Rubin, ``Matching to remove bias in observational studies,''
  \emph{Biometrics}, vol.~29, no.~1, pp. 159--183, 1973.

\bibitem{bib:HoImaiEtAl11m}
\BIBentryALTinterwordspacing
D.~Ho, K.~Imai, G.~King, and E.~A. Stuart, ``Matchit: Nonparametric
  preprocessing for parametric causal inference,'' \emph{Journal of Statistical
  Software, Articles}, vol.~42, no.~8, pp. 1--28, 2011. [Online]. Available:
  \url{https://www.jstatsoft.org/v042/i08}
\BIBentrySTDinterwordspacing

\bibitem{bib:EybenSchererEtAl16gg}
F.~Eyben, K.~R. Scherer, B.~W. Schuller, J.~Sundberg, E.~Andr{\'e}, C.~Busso,
  L.~Y. Devillers, J.~Epps, P.~Laukka, S.~S. Narayanan \emph{et~al.}, ``The
  {Geneva} minimalistic acoustic parameter set {GeMAPS} for voice research and
  affective computing,'' \emph{{IEEE} Trans. Affect. Comput.}, vol.~7, no.~2,
  pp. 190--202, 2016.

\bibitem{bib:MacWhinney17t}
\BIBentryALTinterwordspacing
B.~MacWhinney, ``Tools for analyzing talk part 2: The {CLAN} program,'' 2017,
  pittsburgh, PA: Carnegie Mellon University. [Online]. Available:
  \url{http://talkbank.org/manuals/CLAN.pdf}
\BIBentrySTDinterwordspacing

\bibitem{covington2010cutting}
M.~A. Covington and J.~D. McFall, ``Cutting the gordian knot: The
  moving-average type--token ratio (mattr),'' \emph{Journal of quantitative
  linguistics}, vol.~17, no.~2, pp. 94--100, 2010.

\bibitem{bib:MATLAB19}
MATLAB, \emph{version 9.6 (R2019a)}.\hskip 1em plus 0.5em minus 0.4em\relax
  Natick, Massachusetts: The MathWorks Inc., 2019.

\end{thebibliography}

\end{document}